\documentclass[12pt]{article}
\usepackage[dvips]{color}
\usepackage{epsfig}
\usepackage{amsmath}
\usepackage{graphicx}

\begin{document}
\title{\bf Effects of Thermal Fluctuations on the Thermodynamics of Modified Hayward Black Hole}
\author{{Behnam Pourhassan\thanks{Email:
b.pourhassan@du.ac.ir}\hspace{1mm}, Mir Faizal$^{b}$\thanks{Email:
f2mir@uwaterloo.ca}\hspace{1mm} and Ujjal
Debnath$^{c}$\thanks{Email: ujjaldebnath@gmail.com ;
ujjal@iucaa.ernet.in}}\\
{\small {\em  School of Physics, Damghan University, Damghan, Iran}}\\
$^{b}${\small {\em Deptartment of Physics and Astronomy, University of Lethbridge,  Lethbridge, AB T1K 3M4, Canada}}\\
$^{c}${\small {\em Department of Mathematics, Indian Institute of Engineering Science and Technology,}}\\
{\small {\em Shibpur, Howrah-711 103, India.}}}
\date{}
\maketitle
\begin{abstract}
In this work, we analyze the effects of thermal fluctuations on
the  thermodynamics of a modified Hayward black hole. These thermal
fluctuations will produce correction terms for various
thermodynamic quantities like entropy, pressure, inner energy and
specific heats. We also investigate the effect of these correction
terms on the first law of thermodynamics. Finally, we study the
phase transition for  the  modified Hayward black hole. It is
demonstrated that  the  modified Hayward black hole is stable even
after the thermal fluctuations are taken into account, as long as
the event horizon is larger than a certain critical value.
\\\\
\noindent {\bf Keywords:} Black hole; Thermodynamics.
\end{abstract}

\section{Introduction}
The black hole entropy obtained by famous formula $S = A/4$, where
$A$ denotes the area of the event horizon \cite{4a}. This is the
maximum entropy contained by any object of the same volume
\cite{1, 1a}. It may be noted that a maximum entropy has to be
associated with the black holes to prevent the violation of the
second law of thermodynamics \cite{2, 4}. The reason is that, if a
black hole did not have any entropy, then the entropy of the
universe would spontaneous reduce when any object crosses the
horizon. The observation that the entropy of a black hole scales
with its  area has  led to the development  of the  holographic
principle~\cite{5, 5a}. This principle equates the degrees of
freedom in a region of space with the degrees of freedom on the
boundary of that region.\\

Even though the holographic principle is expected to hold for
large regions of space, it is expected to get violated near the
Planck scale \cite{6, 6a}. This violation of the holographic
principle occurs due to the quantum fluctuations in the geometry
of space-time. As these quantum fluctuations are expected to
dominate the geometry of space-time near the Planck scale, it is
expected that the holographic principle will be violated near the
Planck scale. Thus, the relation between the area and entropy of a
black hole is also expected to get modified near the Planck scale.
In fact, the quantum fluctuations in the geometry of the black
hole will lead to the thermal fluctuations in the black hole
thermodynamics \cite{l1, SPR}. It will be possible to neglect
these thermal fluctuations for the large black holes. However, as
the black holes reduce in size due to the radiation of the Hawking
radiation, the quantum fluctuations in the geometry of the black
hole will increase. Thus, the thermal fluctuations will start to
modify the thermodynamics of the black holes as the black holes
reduce in size. It is possible to calculate the correction terms
generated from such thermal fluctuations. The correction terms
generated from these thermal fluctuations are logarithmic functions
of the original thermodynamic quantities.\\

The corrections to the black hole thermodynamics have   been
obtained using the density of microstates for asymptotically flat
black holes \cite{1z}. This was done by using a formalism called
the   non-perturbative quantum   general relativity.  In this
formalism, the density of states for a black hole was associated
with the  conformal blocks of a well defined conformal field
theory. It was demonstrated that even though the leading order
relation between the entropy and area of a black hole is the
standard Bekenstein entropy-area relation, this formalism also
generated logarithmic corrections terms to the standard Bekenstein
entropy-area relation. It has also been demonstrated using the
Cardy formula that the logarithmic correction terms are generated
for all black holes whose microscopic degrees of freedom are
described by a conformal field theory \cite{card}. Matter fields
have been studied in the presence of a  black hole, and this
analysis has also generated logarithmic correction terms for the
Bekenstein entropy-area formula \cite{other, other0, other1}. The
logarithmic correction terms are also generated from string
theoretical effects \cite{solo1, solo2, solo4, solo5}. The
corrections term for the entropy of a dilatonic black holes has
been calculated \cite{jy}. It was found that this correction term
is again a logarithmic functions of the original thermodynamic
quantities. Such correction terms have also been generated using a
Rademacher expansion of the partition function \cite{bss}. So, it
seems that such  logarithmic corrections terms
occur almost universally.\\

The singularity at the center of black holes indicates a breakdown
of general theory of relativity, as it cannot be gauged away by
coordinate transformations. However, it is possible to construct
black holes which are regular and do not contain an singularity at
the center. The Hayward black hole is an example of such regular
black hole as it does not contain a singularity at the center
\cite{Hayward, HOM}. These black holes have been analyzed using
various modifications of the Chaplygin gas formalism \cite{CG,
GCG, MCG, ECG1,ECG2,ECG3,ECG4}. The motion of a particle in
background of a Hayward black hole also has been discussed
\cite{AS}. The massive scalar quasinormal modes of the Hayward
black hole have been studied \cite{Quasinormal}. In fact, the
one-loop quantum corrections to the Newton potential for these
regular black holes has been calculated in \cite{MHBH}. Recently,
the accretion of fluid flow around the modified Hayward black hole
have been analyzed \cite{UD}. The acceleration of particles in
presence of a rotating modified Hayward black hole has also been
investigated \cite{BU}. In this paper, we will analyze the effects
of thermal fluctuations on thermodynamics of a modified Hayward
black hole.

\section{Modified Hayward Black Hole}

The most general spherically symmetric, static line element
describing modified Hayward black hole can be written as
\cite{MHBH,UD},
\begin{equation}\label{c1}
ds^{2}=-f(r)B(r)dt^{2}+\frac{dr^{2}}{B(r)}+r^{2}d\Omega^{2},
\end{equation}
where
\begin{equation}\label{c2}
d\Omega^{2}=d\theta^{2}+\sin^{2}\theta d\phi^{2},
\end{equation}
with
\begin{eqnarray}\label{c3}
B(r)&=&1-\frac{2Mr^{2}}{r^{3}+2Ml^2},\nonumber\\
f(r)&=&1-\frac{\alpha\beta M}{\alpha r^{3}+\beta M},
\end{eqnarray}
where $M$ is the black hole mass, $l$ is the Hubble length which
is related to the cosmological constant, $\alpha$ is a positive
constants and $\beta$ is related with the cosmological constant
\cite{LML}. In the Ref. \cite{UD} it is found that $\alpha>1$. We
will find lower bounds for both $\alpha$ and $\beta$ using the
thermodynamics description.\\
The horizon radius of the black hole can be found by the real
positive root of the following equation,
\begin{equation}\label{c4}
r_{+}^{3}-2Mr_{+}^{2}+2Ml^{2}=0.
\end{equation}
So, one can obtain the black hole mass in terms of the horizon
radius as follows,
\begin{equation}\label{c5}
M=\frac{r_{+}^{3}}{2(r_{+}^{2}-l^{2})},
\end{equation}
which implies that $r_{+}^{2}>l^{2}$. For simplicity we set $l=1$.
Then, the equation (\ref{c4}) has a solution as follow,
\begin{equation}\label{c6}
r_{+}=\frac{X}{3}+\frac{4M^{2}}{3X}+\frac{2}{3}M,
\end{equation}
where,
\begin{equation}\label{c7}
X\equiv\left(8M^{3}-27M+3\sqrt{81M^{2}-48M^{4}}\right)^{\frac{1}{3}},
\end{equation}
which gives an upper bound for the black hole mass,
\begin{equation}\label{c8}
M\leq\frac{3\sqrt{3}}{4}.
\end{equation}
For the special case of $M=\frac{3\sqrt{3}}{4}$, where $X=2M$, the event horizon radius reduces to
$r_{+}=2M=\frac{3\sqrt{3}}{2}$.\\
An important thermodynamic quantity is the entropy which is
related to the black hole horizon area,
\begin{equation}\label{T2}
S_{0}=\pi r_{+}^{2}
\end{equation}
Also, volume of the black hole is given by,
\begin{equation}\label{T3}
V=\frac{4}{3}\pi r_{+}^{3}.
\end{equation}
Temperature of modified Hayward black hole can be written as,
\begin{equation}\label{T1}
T=\left[\frac{1}{4\pi}\sqrt{-g^{tt}g^{rr}}~\frac{d}{dr}g_{tt}
\right]_{r=r_{+}}=\frac{1}{4\pi}\frac{r_{+}^{3}-4M}{2Mr_{+}^{3}}\sqrt{1-\frac{\alpha\beta
M}{\alpha r_{+}^{3}+\beta M}},
\end{equation}
where the black hole mass $M$ is given by the equation (\ref{c5})
and black hole horizon radius is given by the equation (\ref{c6}).
Therefore, we can investigate thermodynamics of black hole in
terms of either black hole mass or radius associated with the
event horizon. Now, the temperature of the black hole can be
simplified to the another form,
\begin{equation}\label{T1-1}
T=\frac{r_{+}^{2}-3}{4\pi r_{+}^{3}}\sqrt{\frac{2\alpha r_{+}^{2}-\alpha\beta-2\alpha+\beta}{2\alpha r_{+}^{2}
-2\alpha+\beta}}.
\end{equation}
There are two conditions to have real positive temperature,
\begin{eqnarray}\label{T1-1-c1}
r_{+}^{2}&\geq&3,\nonumber\\
r_{+}^{2}&\geq&\frac{\alpha\beta+2\alpha-\beta}{2\alpha}.
\end{eqnarray}
Both conditions satisfied simultaneously if we have,
\begin{equation}\label{T1-1-c2}
\beta(1-\frac{1}{\alpha})\geq4.
\end{equation}
It shows that $\alpha>1$ is necessary to have positive $\beta$, as
illustrated by Ref. \cite{UD}. If we assume integer values for
$\alpha$ and $\beta$, quickly we find $\beta\geq8$. Therefore,
$\alpha=2$ and $\beta=8$ are corresponding to the zero-temperature
limit, and both conditions of (\ref{T1-1-c1}) are the same. The
case of $\beta=0$ yields to ordinary Hayward black hole
\cite{Hayward} with the temperature given by,
\begin{equation}\label{T1-2}
T=\frac{r_{+}^{2}-3}{4\pi r_{+}^{3}},
\end{equation}
which reduces to $T=1/4\pi r_{+}$ for the large values of horizon
radius (asymptotic behavior). The pressure can be obtained by
\begin{equation}
P=\frac{T}{2r_{+}}.
\end{equation}
It is interesting to investigate the first law of thermodynamic
\cite{JJP,Dolan2},
\begin{equation}\label{T4}
dM=TdS+VdP+...,
\end{equation}
other terms are corresponding to the black hole rotation and
charge which are absent in our model. It is easy to check that the
equation (\ref{T4}) violates. In order for thermodynamic
quantities satisfy above relation, we have two solutions: the
first is to add rotation or charge to the black hole, the second
is consideration of logarithmic correction.

\section{Logarithmic correction}

\begin{figure}[th]
\begin{center}
\includegraphics[scale=.5]{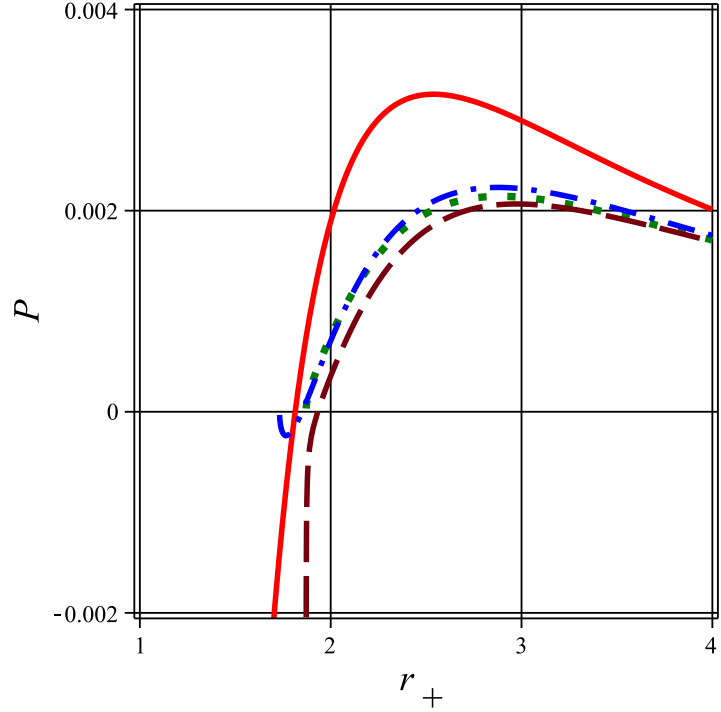}
\caption{Pressure versus horizon radius with $\alpha=2$. $b=1$, $\beta=10$ (dashed line); $b=1$, $\beta=0$
(solid line); $b=0$, $\beta=10$ (dotted line); $b=1$, $\beta=8$ (dash dotted line).}
\label{fig1}
\end{center}
\end{figure}

It is possible to calculate the effect of thermal fluctuations on
the thermodynamics of modified Hayward black hole. One can write
the partition function of the system as
\begin{equation}
Z = \int D g  D A e^{- I},
\end{equation}
where $I \to -i I$ is the Euclidean action for this system \cite{Mir1}.
It is possible to relate it to the partition function in the  statistical mechanical as
\begin{equation}
Z = \int_0^\infty  dE \, \,  \rho (E) e^{-\beta_{\kappa} E},
\end{equation}
where $\beta_{\kappa}$ is the inverse of the temperature.
The partition function can be used to calculate the density of states
\begin{eqnarray}
\rho (E) = \frac{1}{2 \pi i} \int^{\beta_{0\kappa}+
i\infty}_{\beta_{0\kappa} - i\infty} d \beta_{\kappa} \, \, e
^{S(\beta_{\kappa})} ,
\end{eqnarray}
where
\begin{equation}
S = \beta_{\kappa}  E   + \ln Z.
\end{equation}
The  entropy  around the equilibrium temperature $\beta_{0\kappa}$
can be obtained by neglecting  all the thermal fluctuations.
However, if  thermal fluctuations are taken into account, then
$S(\beta_{\kappa})$,   can be written as
\begin{equation}\label{fluc}
S = S_0 + \frac{1}{2}(\beta_{\kappa} - \beta_{0\kappa})^2 \left(\frac{\partial^2
S(\beta_{\kappa})}{\partial \beta_{\kappa}^2 }\right)_{\beta_{\kappa} = \beta_{0\kappa}}.
\end{equation}
So, the density of states can be written as
\begin{eqnarray}
\rho (E) = \frac{e^{S_0}}{ 2 \pi i}  \int^{\beta_{0\kappa}+
i\infty}_{\beta_{0\kappa} - i\infty}  d \beta_{\kappa} \, \,  \exp \left(
\frac{1}{2} (\beta_{\kappa}- \beta_{0\kappa})^2 \left(\frac{\partial^2
S(\beta_{\kappa})}{\partial \beta_{\kappa}^2 }\right)_{\beta_{\kappa} = \beta_{0\kappa}}   \right).
\end{eqnarray}
Thus, we obtain
\begin{equation}
\rho(E) = \frac{e ^{S_{0}}}{\sqrt{2 \pi }}
\left[\left(\frac{\partial^2 S(\beta_{\kappa})}{\partial \beta_{\kappa}^2
}\right)_{\beta_{\kappa} = \beta_{0\kappa}}\right]^{1/2}.
\end{equation}
So, we can write
\begin{equation}
S = S_0 -\frac{1}{2} \ln \left[\left(\frac{\partial^2
S(\beta_{\kappa})}{\partial \beta_{\kappa}^2 }\right)_{\beta_{\kappa} =
\beta_{0\kappa}}\right]^{1/2}.
\end{equation}
This expression can be simplified using the relation between the
microscopic degrees of freedom of a black hole and a conformal
field theory. This is because using this relation, the entropy can
be assumed to the form, $S = a_{1} \beta_{\kappa}^m + a_{2}
\beta_{\kappa}^{-n }$, where $  a_{1}, a_{2}, m, n $ are positive
constants \cite{card}. This has  an extremum at $\beta_{0\kappa} =
(na_{2}/ma_{1})^{1/ m+n} = T^{-1}$ and so we can expand the
entropy around this extremum
\begin{eqnarray}
S(\beta_{\kappa}) &=& [(n/m)^{m/(m+n)} + (m/n)^{n/(m+n)} ](a_{1}^n a_{2}^m)^{1/(m+n)}
\nonumber \\
&&
+ \frac{1}{2}[(m+n) m^{(n+2)/(m+n)} n^{(m-2)/(m+n)}]
( a_{1}^{n+2}a_{2}^{m-2})^{{1}/(m+n)}\nonumber \\ && \times
(\beta_{\kappa} - \beta_{0\kappa})^2, \label{a2}
\end{eqnarray}
Now we can write
\begin{eqnarray}
S_0 &=& (n/m)^{m/(m+n)} + (m/n)^{n/(m+n)} (a_{1}^n a_{2}^m)^{{1}/(m+n)},
\nonumber \\
\left(\frac{\partial^2 S(\beta_{\kappa})}{\partial
\beta_{\kappa}^2 }\right)_{\beta_{\kappa}  = \beta_{0\kappa}} &=&
(m+n) m^{(n+2)/(m+n)} n^{(m-2)/(m+n)} \nonumber \\ && \times (
a_{1}^{n+2}a_{2}^{m-2}  )^{1/(m+n)}~~.
\end{eqnarray}
It is possible to solve for  $a_{1}, a_{2}$ and express the
entropy as
\begin{eqnarray}
\left(\frac{\partial^2 S(\beta_{\kappa})}{\partial
\beta_{\kappa}^2 }\right)_{\beta_{\kappa}  = \beta_{0\kappa}} =
\mathcal{Y} S_0 T^2~,
\end{eqnarray}
where
\begin{eqnarray}
\mathcal{Y} &=& \left[  \left(\frac{(m+n) m^{(n+2)/(m+n)}
n^{(m-2)/(m+n)}}{(n/m)^{m/(m+n)} + (m/n)^{n/(m+n)} }
 \right)~\right. \nonumber \\  && \left. \times
 \left(\frac{n}{m} \right)^{2/(m+n)}  \right].
\end{eqnarray}
However, the factor $\mathcal{Y}$ can be absorbed using some
redefinition as it does not depend on the black hole parameters
\cite{l1, SPR}. Thus, we can write
\begin{equation}
 \left(\frac{\partial^2 S(\beta_{\kappa})}{\partial \beta_{\kappa}^2 }\right)_{\beta_{\kappa}
 = \beta_{0\kappa}}  = S_0 \beta_{0\kappa}^{-2}.
\end{equation}
So, the   corrected value of the entropy can be written as
\begin{equation}
S=S_{0}-\frac{1}{2}\ln{S_{0}T^{2}},
\end{equation}
We can write a general expression for the entropy as
\begin{equation}\label{L1}
S=S_{0}-\frac{b}{2}\ln{S_{0}T^{2}},
\end{equation}
where we introduced a parameter $b$ by hand to track corrected
terms, so in the limit $b\rightarrow0$, the original results can
be recovered and $b=1$ yields usual corrections \cite{SPR}. By
using the temperature and entropy given by the equations
(\ref{T1}) and (\ref{T2}) respectively, we can obtain the
following corrected entropy,
\begin{equation}\label{L2}
S=\pi r_{+}^{2}-\frac{b}{2}\ln\left[\frac{(r_{+}^{2}-3)^{2}(2\alpha r_{+}^{2}-\alpha\beta-2\alpha+\beta)}{16\pi
r_{+}^{4}(2\alpha r_{+}^{2}-2\alpha+\beta)}\right].
\end{equation}
It is clear that the logarithmic correction reduce the entropy of
the black hole. We can obtain the entropy at zero-temperature
using appropriate choice of $\alpha$ and $\beta$ ($\alpha=2$ and
$\beta=8$),
\begin{equation}\label{L2-1}
S(T=0)=\pi r_{+}^{2}-\frac{b}{2}\ln\sqrt{\frac{(r_{+}^{2}-3)^{3}}{16\pi r_{+}^{4}(r_{+}^{2}+1)}}.
\end{equation}
As we mentioned already, both zero-temperature conditions (\ref{T1-1-c1})
are the same for $\alpha=2$ and $\beta=8$. So, $r_{+}=\sqrt{3}$ is zero-temperature
condition  where entropy (\ref{L2-1}) become infinite. The fact is that at the
zero-temperature limit, thermal fluctuations vanish and we should set $b=0$. It is
clear from the equation (\ref{L1}), so at the zero-temperature limit we have $S(T=0)=3\pi$.\\
We can calculate pressure using the (\ref{T2}), (\ref{T3}),
(\ref{T1}) and following relation,
\begin{equation}\label{L3}
P=T\left(\frac{\partial S}{\partial V}\right)_{V}.
\end{equation}
In the Fig. \ref{fig1}, we can analyze the behavior of pressure
for various values of parameters. Comparing dashed line and dotted
line, we can find that the logarithmic correction  decrease the
pressure. Using the higher values for the black hole mass from the
relation (\ref{c8}), we find zero pressure around the black hole
horizon. Higher values of the mass yields to negative pressure.
For all cases with $M<\frac{3\sqrt{3}}{4}$, we can see positive
pressure, which yields to zero as $M\rightarrow0$, it is clearly
expected that vanished black hole has no thermodynamics pressure.

\begin{figure}[th]
\begin{center}
\includegraphics[scale=.5]{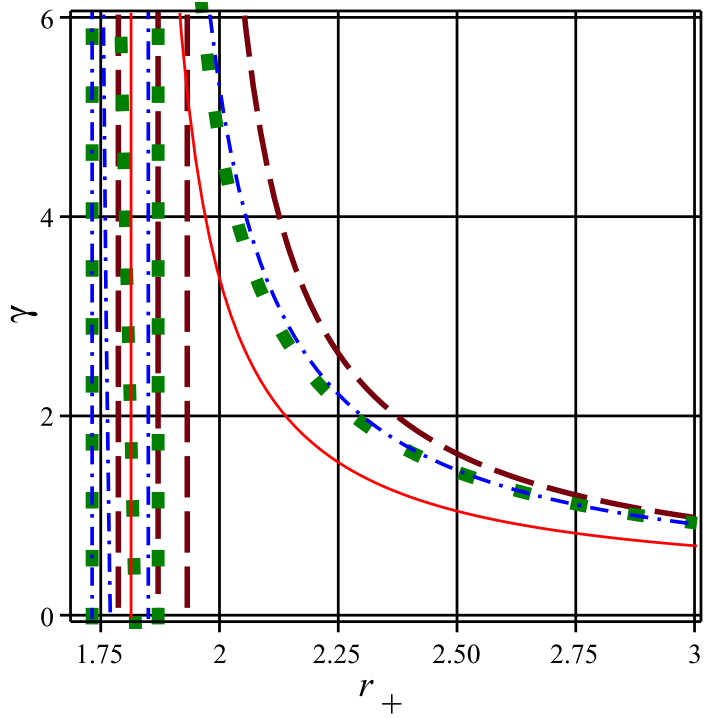}
\caption{$\gamma$ in terms of horizon radius with $\alpha=2$. $b=1$, $\beta=10$ (dashed line); $b=1$, $\beta=0$
(solid line); $b=0$, $\beta=10$ (dotted line); $b=1$, $\beta=8$ (dash dotted line).}
\label{fig2}
\end{center}
\end{figure}

Then, using the well known relation,
\begin{equation}\label{L4}
E=\int{T dS},
\end{equation}
we can obtain inner energy which decreased dramatically due to the logarithmic corrections. It may be noted that
$T$ and $S$ can be used to obtain the specific heat at constant volume,
\begin{equation}\label{L5-0}
C_{v}=\left(T\frac{\partial S}{\partial T}\right)_{V}.
\end{equation}
Calculation of $P$ and $E$ help us to obtain specific heat at constant pressure,
\begin{equation}\label{L5}
C_{p}=\left(\frac{\partial(E+PV)}{\partial T}\right)_{P}.
\end{equation}
So, we can investigate $\gamma=C_{p}/C_{v}$ numerically. From the
Fig. \ref{fig2} it is illustrated that, for the large horizon
radius, $\gamma\rightarrow 0.5$. We find that the value of
$\gamma$ increased due to the logarithmic corrected entropy.

\begin{figure}[th]
\begin{center}
\includegraphics[scale=.5]{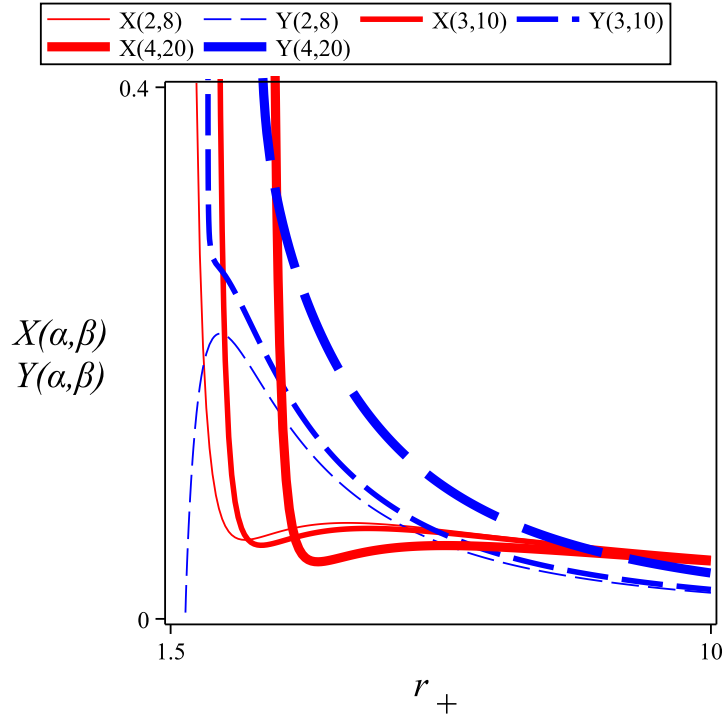}
\caption{X (solid red) and $Y$ (dashed blue) in terms of $r_{+}$
with $b=1$ for various values of $\alpha=2,3,4$ and
$\beta=8,10,20$.} \label{fig3}
\end{center}
\end{figure}

Now, in order to investigate the first law of thermodynamic we
rewrite the equation (\ref{T4}) as follow,
\begin{equation}\label{L6}
X=Y,
\end{equation}
where $X\equiv dM-TdS$ and $Y\equiv VdP$. Then, we give plots of
$X$ and $Y$ in terms of the radius of a black hole horizon in Fig.
\ref{fig3}. We draw three curves corresponding to each
$X(\alpha,\beta)$ and $Y(\alpha,\beta)$. We see that, at least,
there are two points where the first law of thermodynamic is
satisfied. Cross points of the red (solid) and blue (dashed)
curves states $X=Y$ which means validity of the relation
(\ref{T4}). Therefore, we can say that, the first law of
thermodynamic may also valid for the modified Hayward black hole.
There are some special cases with suitable horizon radius where
the first law of thermodynamics is satisfied.

\section{Phase transition}

One of the best ways to find instability of a black hole is
studied the sign of the specific heat given by the equation
(\ref{L5-0}). The black hole is stable for $C_{v}>0$, while is
unstable for $C_{v}<0$. Therefore, $C_{v}=0$ show the point of
phase transition. It is easy to write the specific heat at
constant volume as follow,
\begin{equation}\label{P1}
C_{v}=\frac{A_{1}r_{+}^{8}+A_{2}r_{+}^{6}+A_{3}r_{+}^{4}+A_{4}r_{+}^{2}
+A_{5}}{A_{6}r_{+}^{6}+A_{7}r_{+}^{4}+A_{8}r_{+}^{2}+A_{9}},
\end{equation}
where,
\begin{eqnarray}\label{P2}
A_{1}&=&8\pi \alpha^{2},\nonumber\\
A_{2}&=&4\pi\alpha(2\beta-10\alpha-\alpha\beta),\nonumber\\
A_{3}&=&16\pi\beta\alpha^{2}-2\pi\alpha\beta^{2}-2b\beta\alpha^{2}+56\pi\alpha^{2}
-32\pi\alpha\beta+2\pi\beta^{2}-24b\alpha^{2},\nonumber\\
A_{4}&=&6\pi\alpha\beta^{2}-12\pi\beta\alpha^{2}+18b\beta\alpha^{2}
-24\pi\alpha^{2}+24\pi\alpha\beta-6\pi\beta^{2}+48b\alpha^{2}-24b\alpha\beta,\nonumber\\
A_{5}&=&6b\alpha\beta^{2}-12b\beta\alpha^{2}-24b\alpha^{2}+24b\alpha\beta-6b\beta^{2},\nonumber\\
A_{6}&=&-4\alpha^{2},\nonumber\\
A_{7}&=&4\alpha(\alpha\beta+11\alpha-\beta),\nonumber\\
A_{8}&=&\alpha\beta^{2}-26\beta\alpha^{2}-76\alpha^{2}+40\alpha\beta-\beta^{2},\nonumber\\
A_{9}&=&18\beta\alpha^{2}-9\alpha\beta^{2}+36\alpha^{2}-36\alpha\beta+9\beta^{2}.
\end{eqnarray}

\begin{figure}[th]
\begin{center}
\includegraphics[scale=.5]{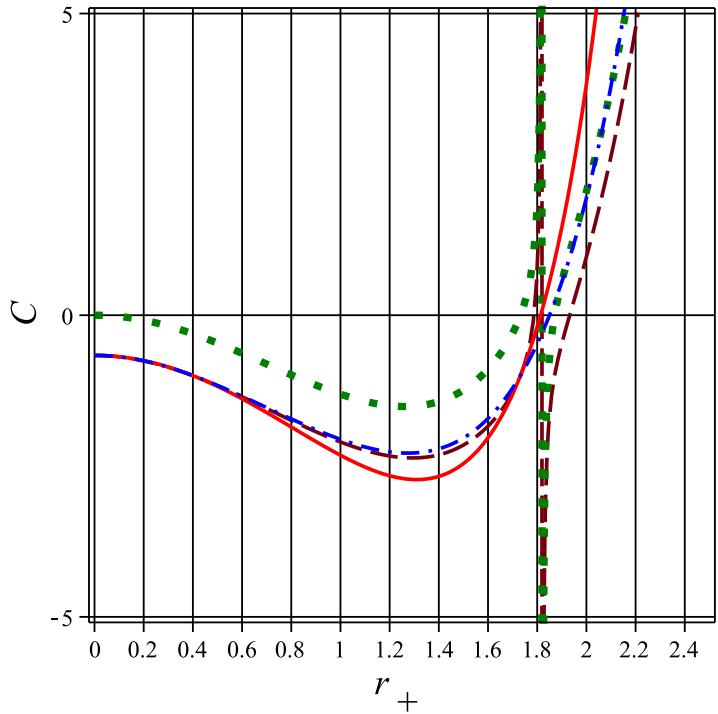}
\caption{Specific heat at constant volume in terms of horizon radius with $\alpha=2$. $b=1$, $\beta=10$
(dashed line); $b=1$, $\beta=0$ (solid line); $b=0$, $\beta=10$ (dotted line); $b=1$, $\beta=8$ (dash dotted line).}
\label{fig4}
\end{center}
\end{figure}

We give graphical analysis of the specific heat which illustrated
in Fig. \ref{fig4}. We can demonstrate that  various cases of the
black hole may be stable for the case of $r_{+}\geq \sqrt{3}$. It
is clear that the modified Hayward black hole with logarithmic
correction of entropy is stable for $r_{+}\geq1.9$. Therefore, we
can say that the modified Hayward black hole is stable for
$r_{+}\geq r_{c}$, where $r_{c}$ is critical value (minimum value)
for the radius of the event horizon. It may be related to the
minimum mass required for the formation of the Hayward black hole
\cite{TSK}. At the zero-temperature limit, where logarithmic
correction vanishes we have $r_{+}=r_{c}=\sqrt{3}$

\section{Conclusions}

In this work, we have studied the spherically symmetric, static
modified Hayward black hole. The entropy, temperature and pressure
have been calculated and found some restrictions on the parameters
$\alpha$ and $\beta$. Next, we have analyzed the effects of
thermal fluctuations on the thermodynamics of a Hayward black
hole. Using the zero-temperature limit of the black hole, we have
obtained the bounds $\alpha\geq2$ and $\beta\geq8$. Here, the
equality holds only in the zero-temperature limit. We have also
analyzed the logarithmic correction to entropy and obtained the
behaviors of the pressure and specific heats numerically. We found
that the value of the pressure and inner energy reduced due to
these logarithmic corrections. We also studied phase transition
for the modified Hayward black hole, and obtained critical point
for such a phase transition. We also demonstrated that the first
law of thermodynamics is satisfied for the modified Hayward black
hole, even in presence of thermal fluctuations. It may be noted
that the thermodynamics of black holes also gets modified because
of the generalized uncertainty principle \cite{mi, r1}. Such
correction terms are non-trivial, and can lead to interesting
consequences like the existence of black hole remnants. It would
be interesting to analyze the corrections to the thermodynamics of
a regular Hayward black hole from the generalized uncertainty
principle.


\begin{thebibliography}{11}
\bibitem{4a}N. Altamirano, D. Kubiznak, R. B. Mann, and Z. Sherkatghanad, Galaxies 2, 89 (2014)
\bibitem{1}J. D. Bekenstein, Phys. Rev. D 7 , 2333 (1973)
\bibitem{1a}J. D. Bekenstein,  Phys. Rev. D 9, 3292 (1974)
\bibitem{2}] S. W. Hawking, Nature 248, 30 (1974)
\bibitem{4}S. W. Hawking, Commun. Math. Phys. 43, 199 (1975)
\bibitem{5} L. Susskind,  J. Math. Phys. 36, 6377 (1995)
\bibitem{5a} R. Bousso,  Rev. Mod. Phys. 74, 825 (2002)
\bibitem{6}  D. Bak and S. J. Rey, Class. Quant. Grav. 17, L1 (2000)
\bibitem{6a}S. K.  Rama, Phys. Lett. B 457, 268 (1999)
 \bibitem{l1}S. Das, P. Majumdar and R. K. Bhaduri,  Class. Quant. Grav. 19, 2355 (2002)
\bibitem{SPR}
J. Sadeghi, B. Pourhassan, and F. Rahimi, Can. J. Phys. 92 (2014) 1638

\bibitem{1z}A. Ashtekar, {  Lectures on Non-perturbative
Canonical Gravity}, World Scientific (1991)
\bibitem{card}T. R. Govindarajan, R. K. Kaul, V. Suneeta, Class. Quant. Grav. 18,  2877 (2001)


\bibitem{other} R. B. Mann and  S. N. Solodukhin, Nucl. Phys. {  B523},  293
(1998)
\bibitem{other0}
A. J. M. Medved and  G. Kunstatter, Phys. Rev. {  D60},   104029 (1999)
\bibitem{other1}
A. J. M. Medved and G. Kunstatter, Phys. Rev. {  D63}, 104005 (2001)
\bibitem{solo1} S. N. Solodukhin, Phys. Rev. {  D57},  2410 (1998)
\bibitem{solo2} A.  Sen,  JHEP  04, 156  (2013)
\bibitem{solo4} A.  Sen,  Entropy  13,  1305 (2011)
\bibitem{solo5} D. A. Lowe and S.  Roy,  Phys. Rev. D82, 063508 (2010)
\bibitem{jy} J. Jing and  M. L Yan, Phys. Rev. {  D63},  24003  (2001)
\bibitem{bss} D. Birmingham and S. Sen, Phys. Rev. {  D63}, 47501 (2001)


\bibitem{Hayward}
S.A. Hayward,   Phys. Rev. Lett. 96 (2006) 031103
\bibitem{HOM}
M. Halilsoy, A. Ovgun, S.H. Mazharimousavi,
Eur. Phys. J. C 74 (2014) 2796
\bibitem{CG}
A. Sen,   Phys. Scr. T117 (2005) 70
\bibitem{GCG}
L. Xu, J. Lu, Y. Wang,    Eur. Phys. J. C 72 (2012) 1883
\bibitem{MCG}
U. Debnath, A. Banerjee, S. Chakraborty, Class. Quant. Grav.  21
(2004) 5609
\bibitem{ECG1}
E.O. Kahya, B. Pourhassan,  Astrophys. Space Sci. 353
(2014) 677
\bibitem{ECG2}
B. Pourhassan, E.O. Kahya,
Results Phys. 4 (2014) 101
\bibitem{ECG3}
B. Pourhassan and E. O. Kahya,    Advances
in High Energy Physics 2014 (2014) 231452
\bibitem{ECG4}
E.O. Kahya, M. Khurshudyan, B. Pourhassan, R. Myrzakulov, A. Pasqua,   Eur. Phys. J. C 75 (2015) 43
\bibitem{AS}
G. Abbas, U. Sabiullah,  Astrophys. Space Sci. 352 (2014) 769
\bibitem{Quasinormal}
K. Lin, J. Li and S. Yang,   Int. J. Theor. Phys. 52
(2013) 3771
\bibitem{MHBH}
T. De Lorenzo, C. Pacilioy, C. Rovelli and S. Speziale,
Gen. Rel. Grav. 47 (2015)  41
\bibitem{UD}
U. Debnath,   Eur. Phys. J. C75 (2015) 3, 129
\bibitem{BU}
B. Pourhassan and  U. Debnath,  [arXiv:1506.03443 [gr-qc]]
\bibitem{LML}
J. Li, H. Ma, K. Lin,
Phys.  Rev.  D 88 (2013) 064001
\bibitem{JJP}
J. Sadeghi, K. Jafarzade, B. Pourhassan,
Int. J. Theor. Phys. 51 (2012) 3891
\bibitem{Dolan2}
B. P. Dolan,  Class. Quant. Grav. 28 (2011)
235017
\bibitem{Mir1}
B. Pourhassan and Mir Faizal,   Europhys. Lett. 111: 40006, 2015
\bibitem{TSK}
R. Tharanath, J. Suresh, V.C. Kuriakose,
Gen. Rel. Grav. 47 (2015) 46
\bibitem{mi}
M. Faizal and M. Khalil,    Int.J.Mod.Phys. A30 (2015) 1550144
\bibitem{r1}
A. F. Ali, JHEP 1209, 067  (2012)

\end{thebibliography}
\end{document}